\documentclass[twocolumn]{aastex63}
\usepackage{xspace,color,soul}

\newcommand{\per}{\ensuremath{^{-1}}\xspace}
\newcommand{\Lya}{Ly\ensuremath{\alpha}\xspace}

\accepted{July 2022}

\shortauthors{Songaila et al.}

\begin{document}

\title{The Evolution of Lyman Alpha Emitter Line Widths from $z$=5.7 to $z$=6.6}

\correspondingauthor{Antoinette Songaila}
\email{acowie@ifa.hawaii.edu}

\author{A.~Songaila}
\affiliation{Institute for Astronomy, University of Hawaii,
2680 Woodlawn Drive, Honolulu, HI 96822, USA}

\author[0000-0002-3306-1606]{A.~J.~Barger}
\affiliation{Department of Astronomy, University of Wisconsin-Madison,
475 N. Charter Street, Madison, WI 53706, USA}
\affiliation{Department of Physics and Astronomy, University of Hawaii,
2505 Correa Road, Honolulu, HI 96822, USA}
\affiliation{Institute for Astronomy, University of Hawaii, 2680 Woodlawn Drive,
Honolulu, HI 96822, USA}

\author[0000-0002-6319-1575]{L.~L.~Cowie}
\affiliation{Institute for Astronomy, University of Hawaii,
2680 Woodlawn Drive, Honolulu, HI 96822, USA}

\author{E.~M.~Hu}
\affiliation{Institute for Astronomy, University of Hawaii,
2680 Woodlawn Drive, Honolulu, HI 96822, USA}

\author[0000-0003-1282-7454]{A.~J.~Taylor}
\affiliation{Department of Astronomy, University of Wisconsin-Madison,
475 N. Charter Street, Madison, WI 53706, USA}


\begin{abstract}
Recent evidence suggests that high-redshift Ly$\alpha$ emitting galaxies (LAEs) 
with $\log L({\rm Ly}\alpha) > 43.5~{\rm erg \ s}^{-1}$, referred to as ultraluminous 
LAEs (ULLAEs), may show less evolution than 
lower-luminosity LAEs in the redshift range $z=5.7$--6.6. Here we explore the redshift evolution of  the 
velocity widths of the Ly$\alpha$ emission lines in LAEs over this redshift interval.  We use new wide-field,
narrowband observations from Subaru/Hyper Suprime-Cam to provide
a sample of 24 $z=6.6$ and 12 $z=5.7$ LAEs with $\log L({\rm Ly}\alpha) >43~{\rm  erg\ s}^{-1}$,
all of which have follow-up spectroscopy from Keck/DEIMOS.  Combining with archival lower-luminosity
data, we find a significant narrowing of the Ly$\alpha$ lines in LAEs at
$\log L({\rm Ly}\alpha) < 43.25~{\rm  erg\ s}^{-1}$---somewhat lower than the usual ULLAE
definition---at $z=6.6$ relative to those at $z=5.7$, but we do not see this in higher-luminosity LAEs.
As we move to higher redshifts, the increasing neutrality of the intergalactic medium 
should increase the scattering of the Ly$\alpha$ lines, making them narrower. The absence of this
effect in the higher-luminosity LAEs suggests they may lie in more highly ionized regions, 
self-shielding from the scattering effects of the intergalactic medium.
\end{abstract}


\section{Introduction}
Ly$\alpha$ emitting galaxies (LAEs) may provide our current strongest
probe of the epoch of reionization, which occurred at 
$z\sim7$. The evolution of LAEs can provide powerful
diagnostics of the physics of the ionization of the intergalactic medium
(IGM), the sources of the ionizing photons, and the structure of the
ionized regions in the IGM. As the neutral hydrogen fraction
in the IGM increases with increasing redshift, we expect that Ly$\alpha$
emission lines should become narrower and less
luminous and that only the red wings of the lines should be seen
(see, e.g., \citealt{hayes21} and references therein).
At $z>5.5$, this is observed for LAEs with Ly$\alpha$ luminosities
$L (\textrm{\Lya}) <10^{43.5}$~erg~s$^{-1}$.  

However, the advent of giant imagers, such
as Subaru/Hyper Suprime-Cam \citep[HSC; ][]{miyazaki18}, has allowed an expansion in
the range of luminosities that can be observed, making it possible to probe 
the rarer ultraluminous LAEs
($L (\textrm{\Lya}) >10^{43.5}$~erg~s$^{-1}$; ULLAEs).
Based on the mapping of these sources, recent work has suggested that
ULLAEs do not show such evolution.

One route for probing the evolution of LAEs at these high redshifts is 
by measuring their luminosity functions (LFs).
\cite{santos16} were the first to claim no evolution in their photometric LAE LFs
at the ultraluminous end after observing that their $z=5.7$ and $z=6.6$ LFs converged near
$\log L (\textrm{\Lya})\approx 43.6$ erg s\per. They interpreted this result as evidence 
that the most luminous LAEs formed ionized bubbles around themselves, thereby 
becoming visible at earlier
redshifts than lower-luminosity LAEs. The normalizations of the \citet{santos16}
LFs have been questioned in subsequent papers \citep{konno18,taylor20}. 
However, recent analyses by \cite{taylor20,taylor21} and \cite{ning22}
are also consistent with no
evolution in the ULLAE LF, though the results are not
highly significant, as we discuss further in the summary.

If the most luminous LAEs do indeed generate ionized bubbles, then
theoretical modeling can be used to infer key properties
of the galaxies, such as the escape fraction of ionizing photons \citep[e.g.,][]{gronke21}.

The formation of ionized bubbles is also suggested by
the discovery of double-peaked spectra in some ULLAEs at $z=6.6$.
While \Lya line profiles at $z\sim3$ show double-peaked spectra (both red and blue
peaks) in $\sim$30\% of cases \citep{kulas12}, at $z>5$, it was expected that 
the blue peak should  always be scattered
away by the neutral portion of the IGM \citep{hu10,hayes21}, 
leaving a single peak featuring a sharp blue break and an
extended red wing. However, \cite{hu16} and \cite{songaila18}
reported $z=6.6$ double-peaked ULLAEs
in the COSMOS (COLA1) and
North Ecliptic Pole (NEP) (NEPLA4) fields, respectively.
More recently, \cite{meyer21} found a $z=6.8$ double-peaked LAE at a luminosity 
of $\log L (\textrm{\Lya})=42.99$~erg~s\per in the A370p field (A370p\_z1),
and \cite{bosman20} found a $z=5.8$ double-peaked LAE
at a luminosity of $\log L (\textrm{\Lya})=43.03$~erg~s\per in a quasar 
proximity zone (Aerith B).
In a theoretical analysis, \cite{gronke21} showed that ionized 
bubbles around such objects can allow the double-peaked structure to be seen.

In the present paper, we investigate another route for probing
the evolution of LAEs at these high redshifts, namely, by mapping the velocity
widths of the Ly$\alpha$ profiles as a function of redshift and
luminosity. We use wide-field narrowband observations
of the NEP made with Subaru/HSC to develop substantial samples
of luminous LAEs, all of which have follow-up spectroscopy from Keck/DEIMOS. 
In combination with lower-luminosity observations from \cite{hu10} (note that, 
for simplicity, we will refer to their 
$z=6.5$ sample as being at $z=6.6$), this
provides a set of  homogeneously observed velocity profiles over
a wide range of luminosities at $z=5.7$ and $z=6.6$.

We assume $\Omega_M=0.3$, $\Omega_\Lambda=0.7$, and
$H_0=70$~km~s$^{-1}$~Mpc$^{-1}$ throughout.
All magnitudes are given in the AB magnitude system,
where an AB magnitude is defined by
$m_{AB}=-2.5\log f_\nu - 48.60$.
Here $f_\nu$ is the flux of the source in units of
ergs~cm$^{-2}$~s$^{-1}$~Hz$^{-1}$.

\section{Data}
\label{secdata}

\subsection{Target selection}
Our LAE sample is primarily drawn from the HEROES survey centered 
on the NEP \citep{songaila18}. The HEROES observations in the optical
were made with Subaru/HSC and 
were reduced with the hscPipe software, which was also
used to generate the catalogs (A. Taylor et al.\ 2022, in preparation). 
For each object, we computed the magnitudes using
$2^{\prime\prime}$ diameter apertures, and we corrected these to total
magnitudes using the median offset between
$2^{\prime\prime}$  and $4^{\prime\prime}$  apertures.
These offsets are typically around
-0.2~mag. The corrected aperture magnitudes
match well to Kron magnitudes measured on the galaxies.
The HEROES data have very high-quality spatial resolution,
about $0.5^{\prime\prime}$--$0.6^{\prime\prime}$ full width at half maximum
(FWHM) for most of the colors throughout most of the field.

The full HSC observations of
HEROES cover $50.2~{\rm deg}^2$ in 5 broadbands:
 $g$: 27.3, $r$: 26.9, $i$: 26.5, $z$: 26.0, and $Y$: 25.3, 
where the numbers are the median $1\sigma$ noise across the field  
in the corrected  $2\arcsec$ diameter apertures.
The field is also imaged in two narrowband filters, NB816 and NB921,
with $1\sigma$ depths of 25.9 and 25.7, respectively.

We also obtained deeper observations around the {\em JWST\/} Time Domain 
Field (JTDF) \citep{jansen18}.
The JTDF is a $14\arcmin$ diameter field that lies within the footprint of HEROES.
It will be intensively observed with {\em JWST\/}, as well as
with {\em HST\/}, {\em Chandra\/}, and ground-based telescopes.
The NB921 observations of this region cover $2.1~{\rm deg}^2$ 
and have a median $1\sigma$ depth of 26.2, or about
0.5~mag deeper than the average sensitivity.

\begin{figure*}
\includegraphics[width=9cm,angle=0]{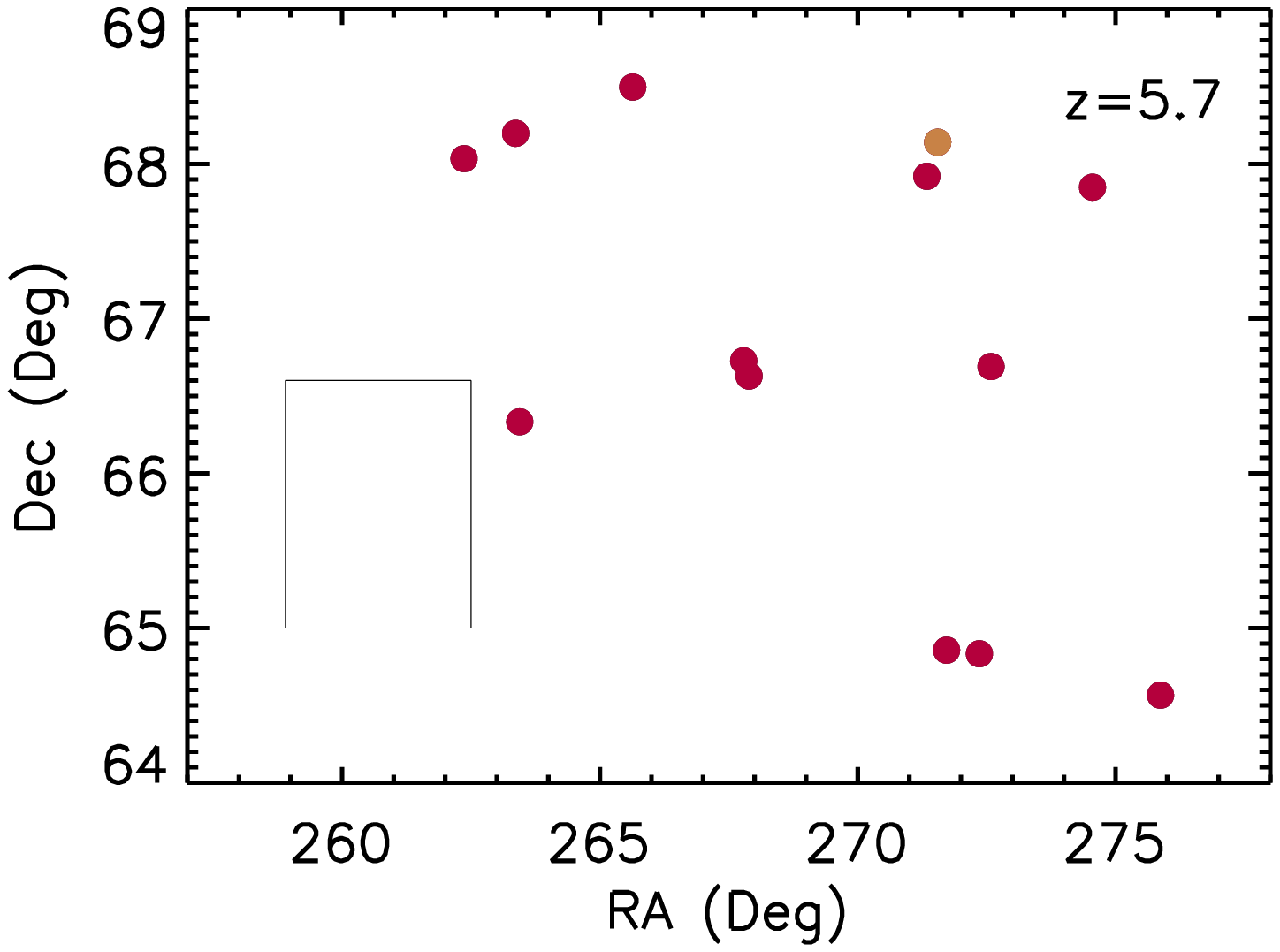}
\includegraphics[width=9cm,angle=0]{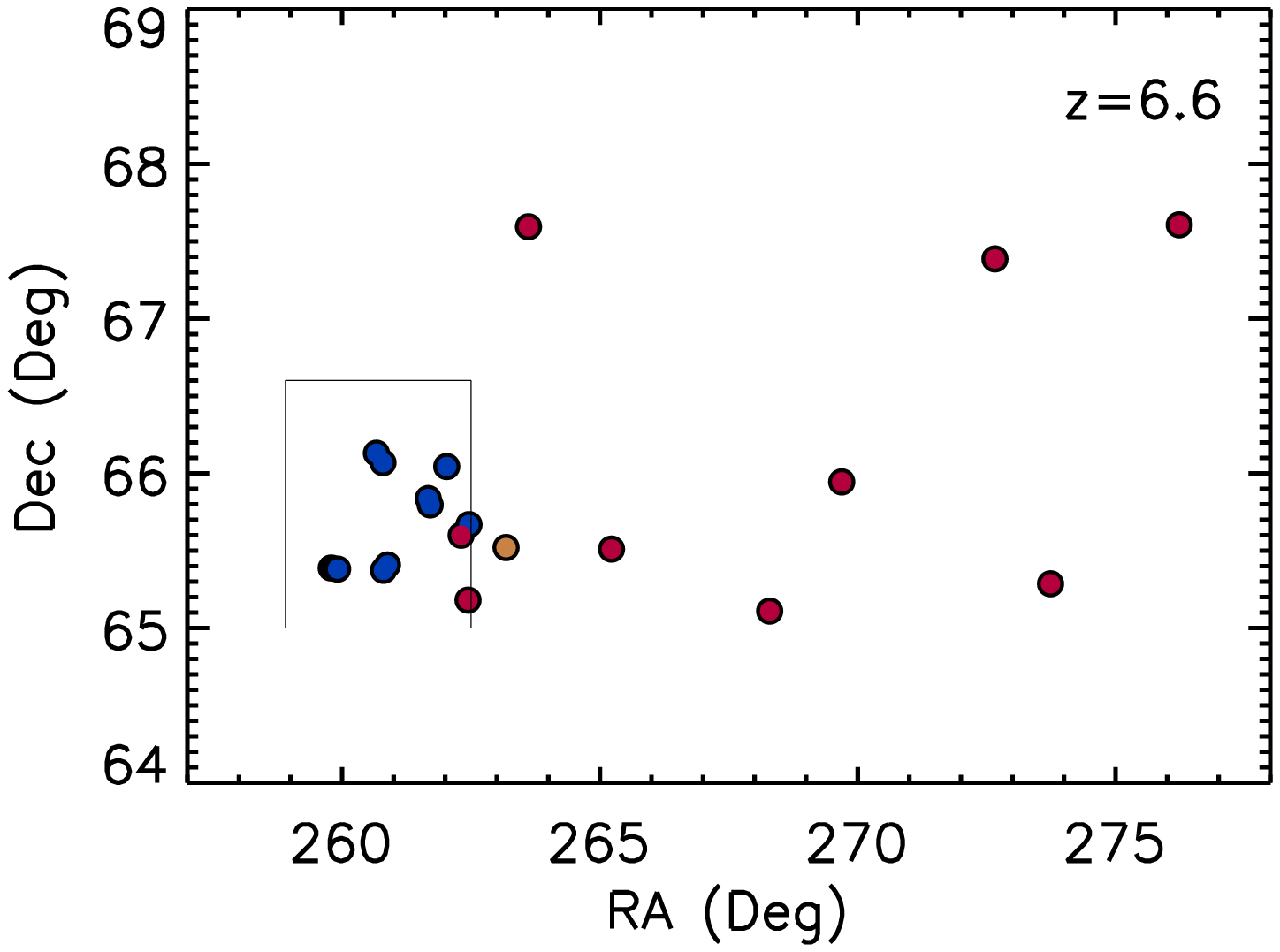}
\caption{{\em (Left)\/} The $z=5.7$\ HEROES LAE sample (Table~\ref{lumtab}; red circles).
{\em (Right)\/} The $z=6.6$\ HEROES (Table~\ref{lumtab2}; red circles) and JTDF  
(Table~\ref{lumtab3}; blue circles) LAE samples.
In both panels, the new Subaru/HSC data surrounding the JTDF field
are marked with a rectangle, and gold circles show
spectroscopically detected AGNs at the given redshifts.
}
\label{positions}
\end{figure*}

The selection of the $z=6.6$ LAEs from the NB921
imaging is described in \cite{taylor20}, and the selection of
the $z=5.7$ LAEs from the NB816 imaging is described in \cite{taylor21}. 
These papers also
describe the computation of the Ly$\alpha$ line luminosities from
the narrowband magnitudes. While the exact conversion
depends on the position of the Ly$\alpha$ line on the filter, and
hence on the redshift, the $5\sigma \log L (\textrm{\Lya})$ limit
is roughly 42.4~erg~s$^{-1}$ at $z=5.7$ and 
43.0 erg~s$^{-1}$ at $z=6.6$. The $5\sigma \log L (\textrm{\Lya})$
limit in the JTDF field is 42.5~erg~s$^{-1}$ at $z=6.6$.

We summarize the $z=5.7$ LAE sample from 
\cite{taylor21} in Table~\ref{lumtab},
the $z=6.6$ LAE sample from \cite{taylor20}
in Table~\ref{lumtab2}, and the new JTDF LAE sample
from this work in Table~\ref{lumtab3}.
The selection of the JTDF LAE sample precisely
follows the methods used in the two previous papers.
In Table~\ref{lumtab2}, we added three sources from the COSMOS field:
COLA1 from \cite{hu16}  and CR7 and MASOSA from \citet{sobral15}
and \citet{matthee15}. 
We also added one source (VR7) from the SSA22 field
\citep{matthee17} and one source (GN-LA1) from the GOODS-N
field \citep{hu10}. These add additional high-luminosity LAEs
where we have high-quality Keck/DEIMOS
spectra obtained in the same configurations as for the LAEs
in the NEP field.

In the tables,
we only include sources whose redshifts have been confirmed by subsequent 
spectroscopy. We give 
the source name, the R.A. and decl., the redshift corresponding to the
peak of the Ly$\alpha$ line, the $\log L (\textrm{\Lya})$, and the FWHM.
Note that we quote the observed luminosity. We have made no attempt 
to correct for the intergalactic transmission, and the
intrinsic luminosity could be higher, possibly by a
factor of two or more (e.g., \citealt{hu10}).
However, estimating this
correction would not be easy, given the possible presence
of ionized bubbles surrounding these objects, and we
do not do so here. We show the locations of the LAEs on the HEROES and
JTDF fields in Figure~\ref{positions}.

In the present paper, we compare the above samples with less luminous samples from \cite{hu10}.
We re-reduced their spectra and included new data that we obtained
subsequently. 

For consistency, we analyze all the samples using the asymmetric 
fitting procedures described in Section~\ref{secanal}.

\begin{figure*}[th]
\includegraphics[width=6cm,angle=0]{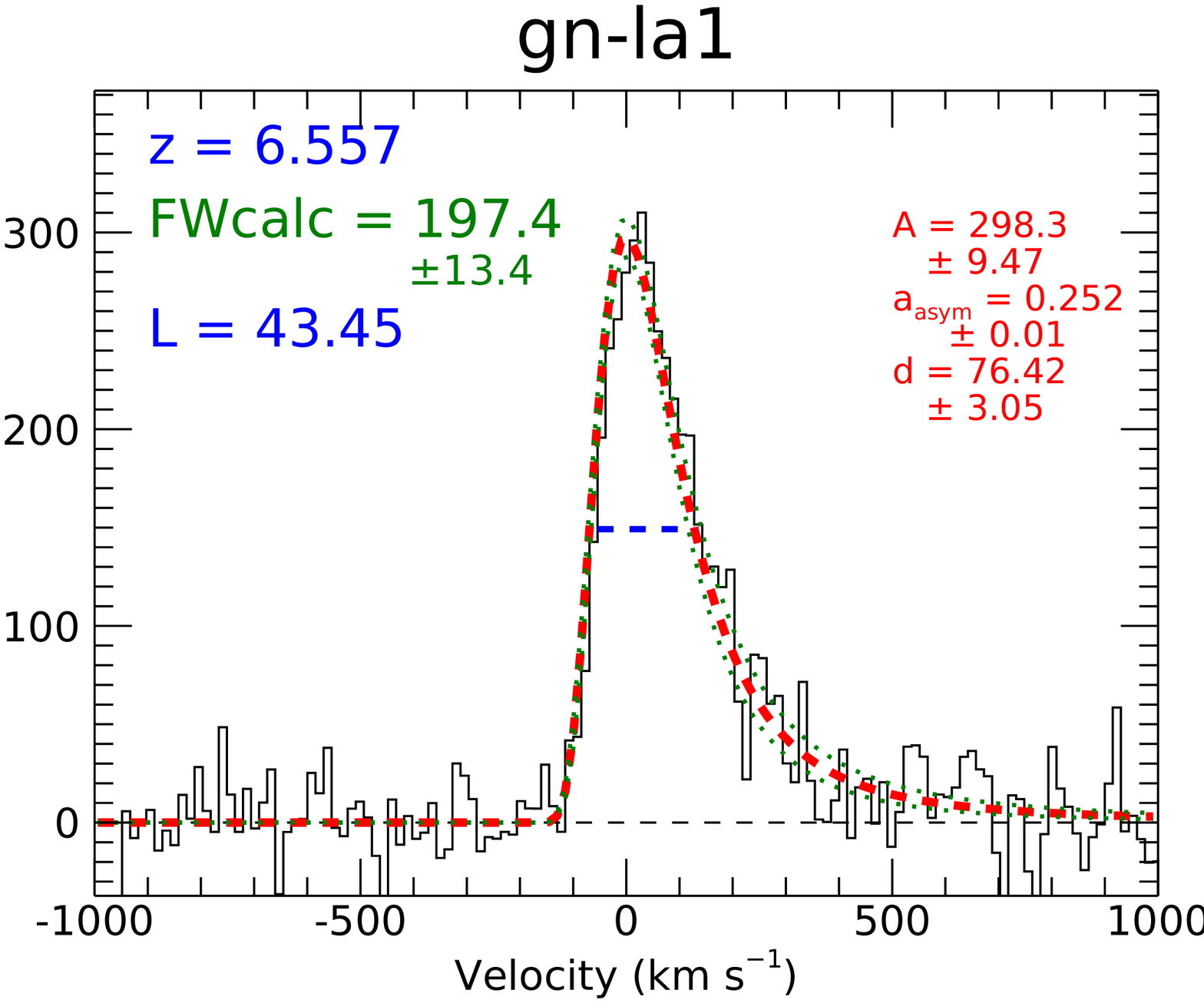}
\includegraphics[width=6cm,angle=0]{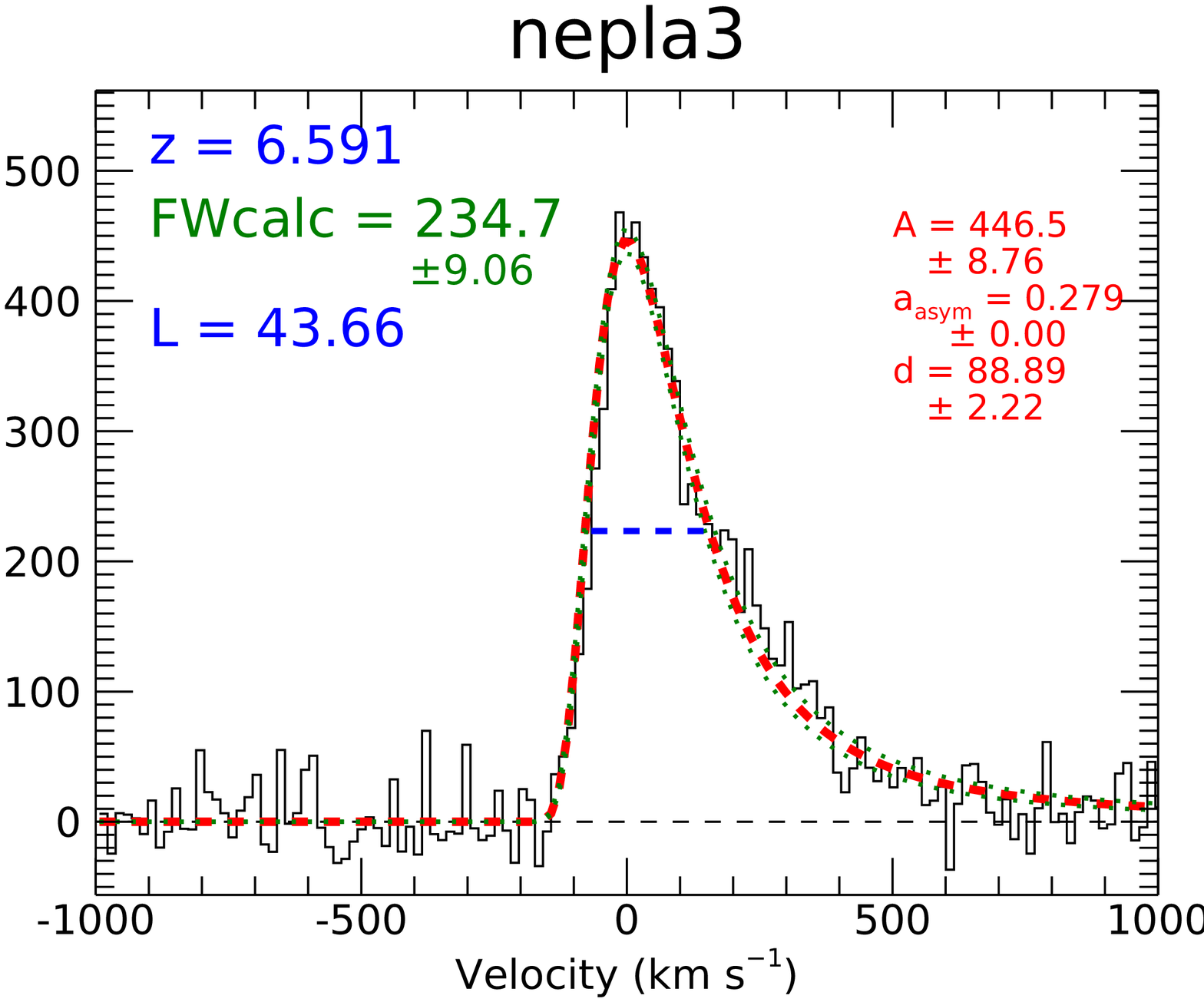}
\includegraphics[width=6cm,angle=0]{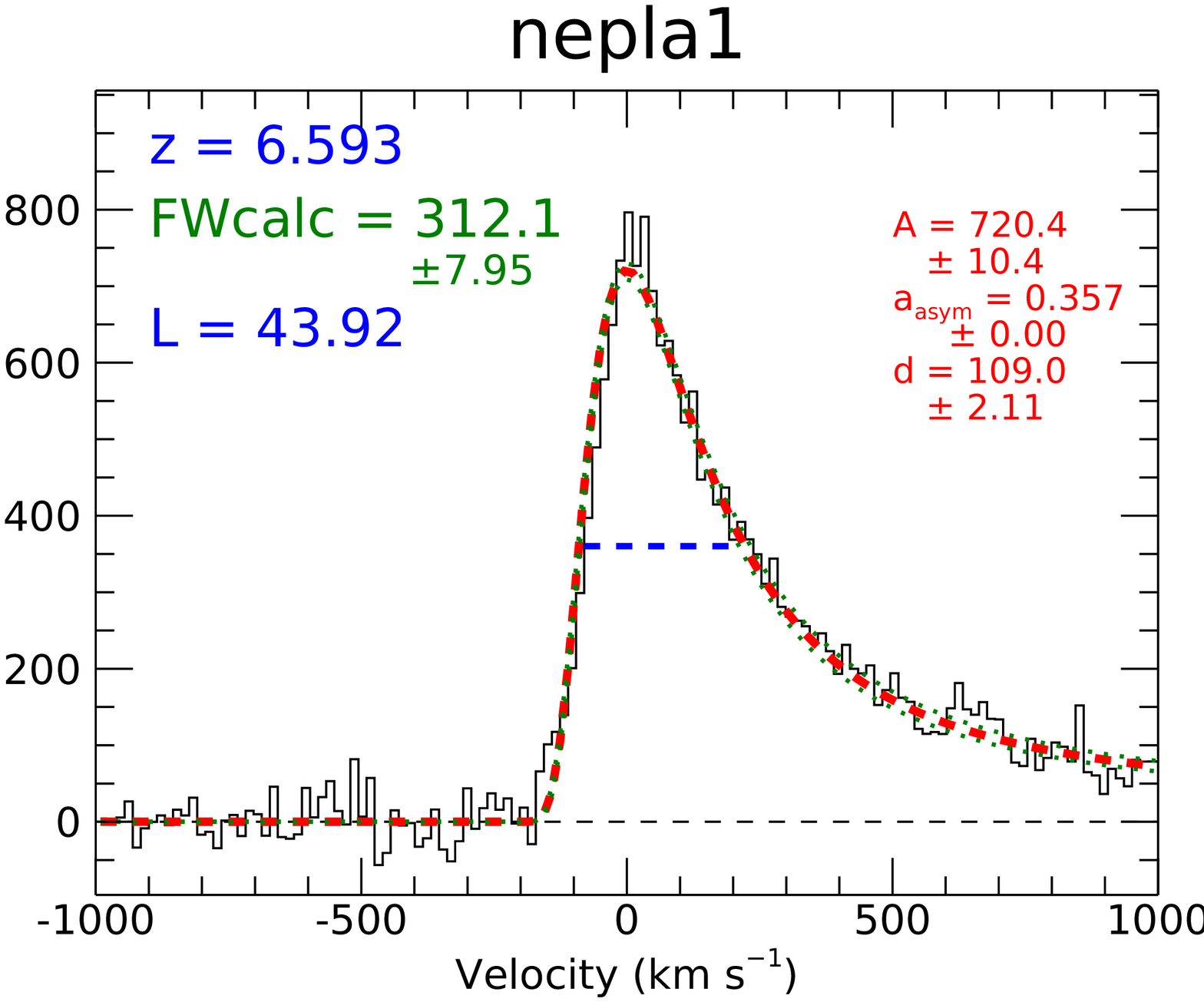}
\includegraphics[width=6cm,angle=0]{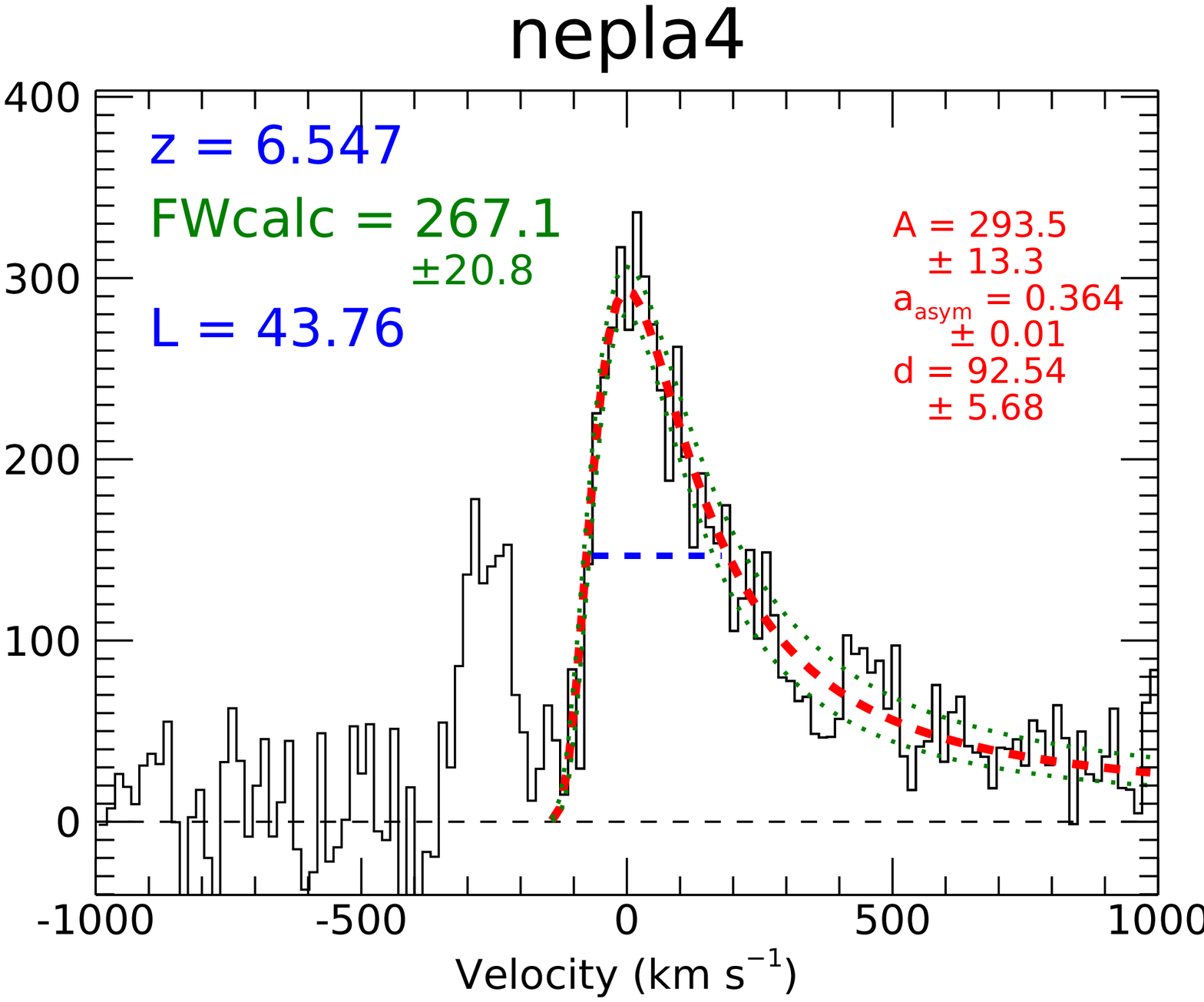}
\includegraphics[width=6cm,angle=0]{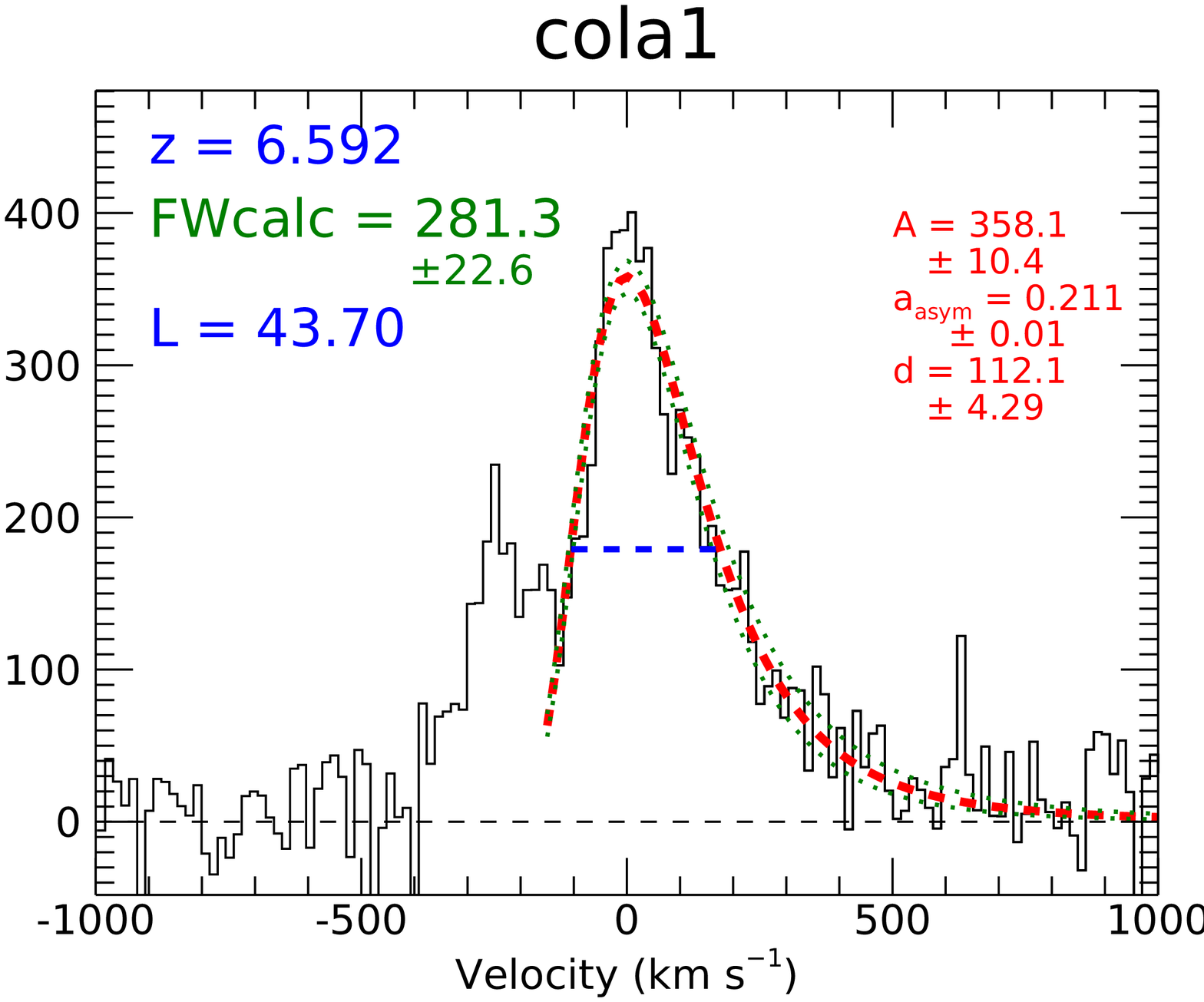}
\caption{Examples of asymmetric fits to $z=6.6$\ LAEs. 
The top three panels show fits to three LAEs with
a range of luminosities. The lower two panels show the two
double-peaked ULLAEs, where we have fitted to only the red
wing. In all cases, the red dashed curve shows the fit
to the data (black), and the green dotted curves show the
$1\sigma$ errors. 
The adopted widths based on the fits, FWcalc in km s$^{-1}$, 
are listed in green and are shown as the blue dashed lines. The redshift and
$\log L (\textrm{\Lya})$ in erg s$^{-1}$ values are 
listed in blue, and the fitting parameters from
Equation~\ref{eq_asymgauss} are listed in red. Because of the 
uncertainties in the spectrophotometric calibration, the 
y-axis shows the flux in arbitrary units.
}
\label{nb921_skew}
\end{figure*}

\begin{figure*}
\includegraphics[width=6cm,angle=0]{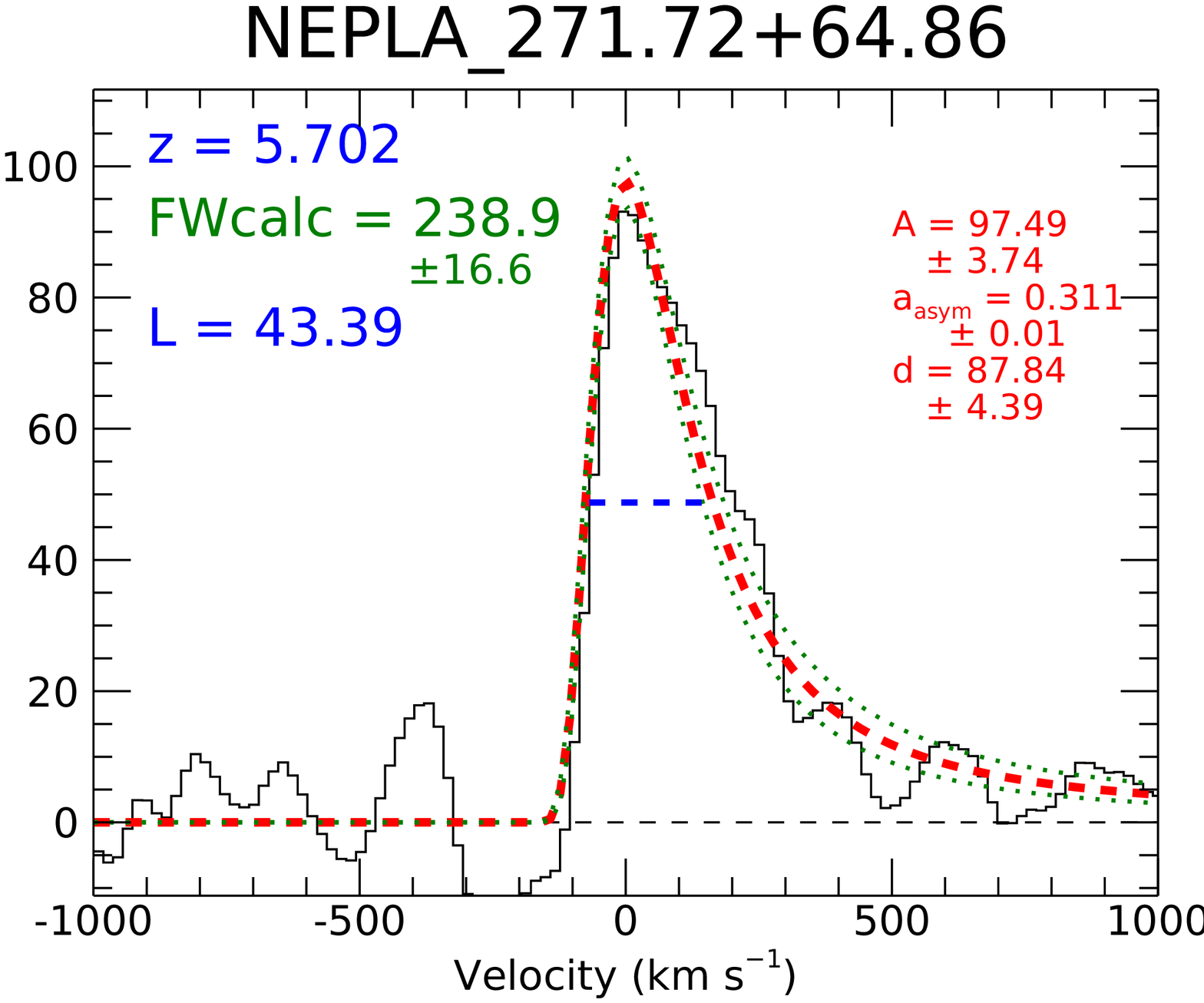}
\includegraphics[width=6cm,angle=0]{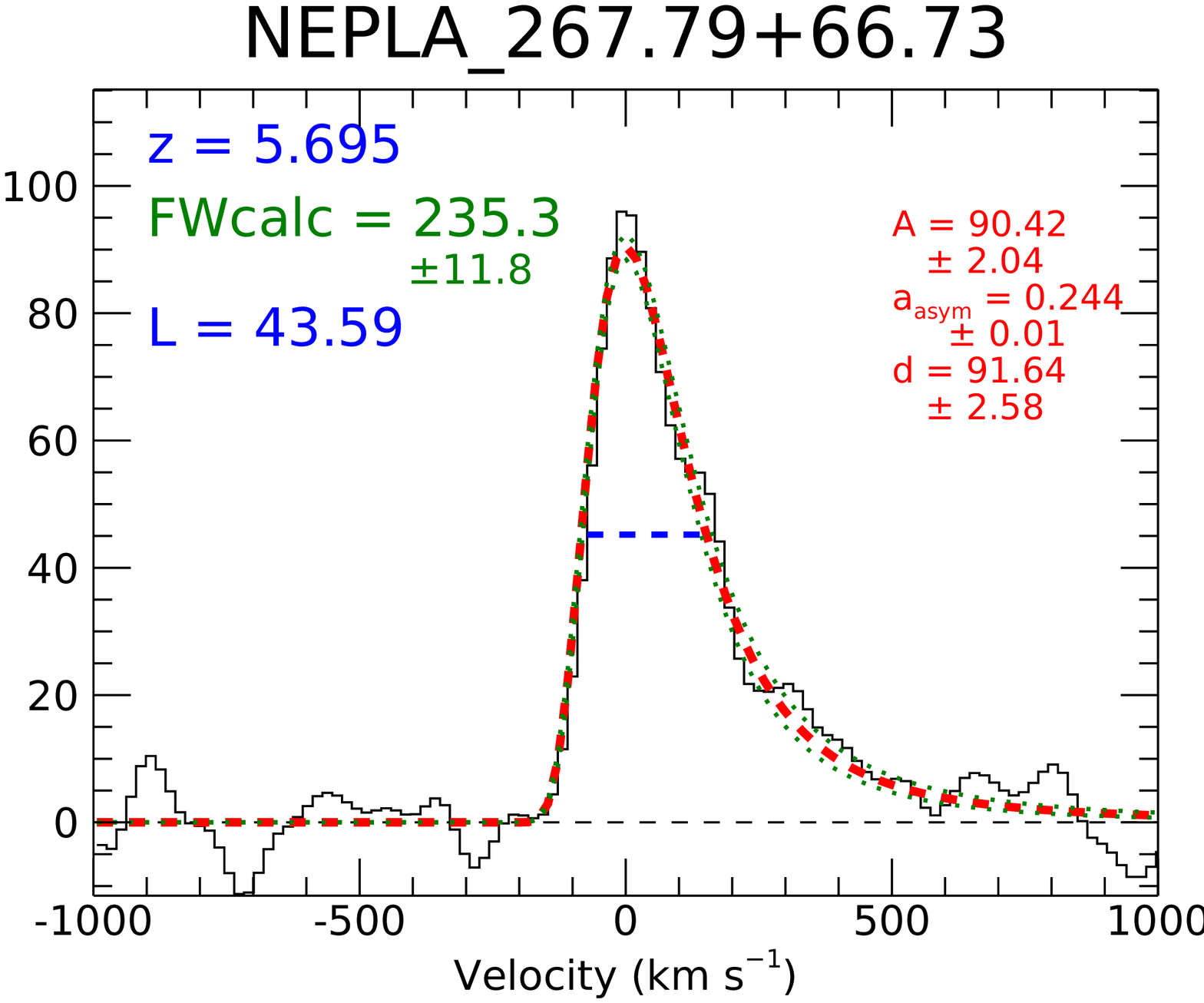}
\includegraphics[width=6cm,angle=0]{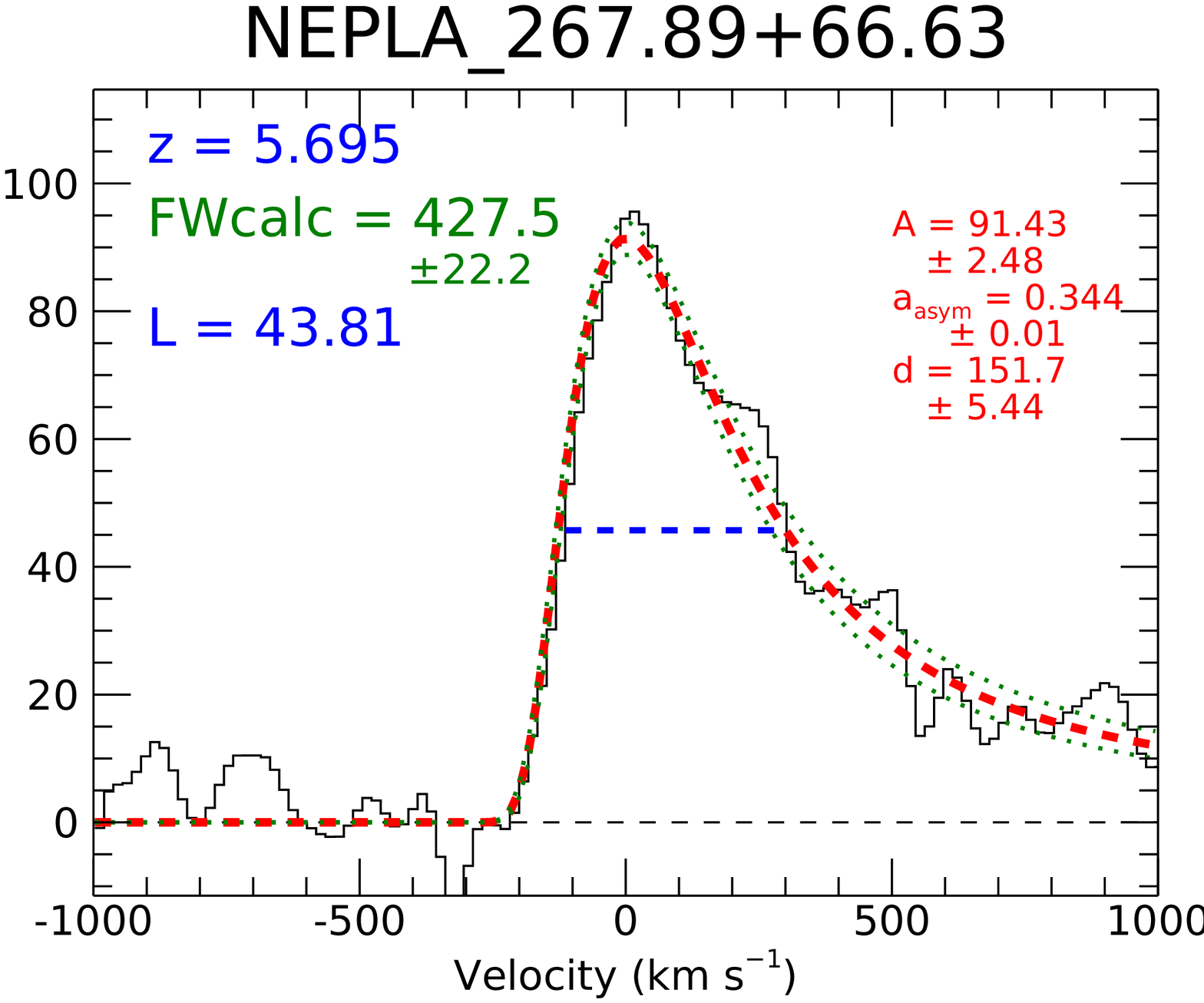}
\caption{Examples of asymmetric fits to $z=5.7$\ LAEs.
The notation is the same as in Figure~\ref{nb921_skew}.
}
\label{nb816_skew}
\end{figure*}

\subsection{Spectroscopy}
We obtained spectroscopic observations with the Keck/DEIMOS spectrograph
for all of the sources in Tables~1, 2, and 3, though we have previously
reported some of them in \cite{songaila18} and in \cite{taylor20,taylor21}.
We configured DEIMOS using $1\arcsec$ slits and the 830G grating, which
has high throughput at 9000~\AA\ and 
provides a resolution of $R=2550$.  We took three 20~min sub-exposures for each slitmask, dithering
$\pm1\farcs5$ along the slits for improved sky subtraction and to minimize CCD systematics.
The minimum total exposure on an individual LAE was 1~hr,
while most were observed for 2--3~hr. GN-LA1 in the intensely observed
GOODS-N field has 14~hr of exposure. 

We reduced the data using the standard pipeline from \cite{cowie96}. We performed an initial 
pixel-by-pixel sky subtraction by combining the three dithered exposures and subtracting the 
minimal value recorded by each pixel. Next, we median combined the three dithered frames, 
adjusting for the $\pm1\farcs5$ offsets. We rejected cosmic rays using a $3\times 3$ pixel 
median rejection spatial filter, and we quantified and corrected for geometric distortions in 
the spectra using pre-selected bright continuum sources from the slitmask. Lastly, we used 
the observed sky lines to calibrate the wavelength scale and to perform a final sky subtraction.

In Figure~\ref{nb921_skew}, we show the spectra of three $z=6.6$ LAEs
with a range of luminosities, along with the two double-peaked ULLAEs.
In Figure~\ref{nb816_skew}, we show the spectra of three $z=5.7$
LAEs with a range of luminosities.
There is no sign of active galactic nucleus (AGN)
activity, such as [NV], in any of the LAE spectra. 
The selection method does pick out a small number of AGNs at these redshifts, 
but they are easily distinguished based on their spectra. There is one AGN in
each sample (see Figure~\ref{positions}).


\section{Line Width Measurements}
\label{secanal}
Following \citet{claeyssens19} and \citet{shibuya14}, we fitted the LAE Ly$\alpha$ lines
with an asymmetric profile,
\begin{equation}\label{eq_asymgauss}
        f(\lambda) = A \exp \left(-\frac{\Delta v^2}{2(a_{\rm asym} (\Delta v) +d)^2}\right) \,,
\end{equation}
where $A$ is the normalization, $\Delta v$\ is the velocity relative to the peak of the Ly$\alpha$ profile,
$a_{\rm asym}$ controls the asymmetry, and $d$ controls the line width.  
We fitted this function to the Ly$\alpha$ lines in the three samples listed in 
Tables~\ref{lumtab}, \ref{lumtab2}, and \ref{lumtab3}
and to the \cite{hu10} sample using the IDL fitting routine MPFIT of \cite{markwardt09}.
In terms of these free parameters, the width of the line is
\begin{equation}\label{eq_asymwidth}
       {\rm FWHM}= \frac{2 \sqrt{2 \ln 2}\  d}{(1-2\ln 2 \ a_{\rm asym}^2)} \,.
\end{equation}

We show the fits for the LAEs in Figures~\ref{nb921_skew}
and \ref{nb816_skew}. In each case, we show the model fit (red)
overlaid on the spectrum (black). We list the fitted parameters from
Equation~\ref{eq_asymgauss} in red in each panel.
For the sources with double peaks, we fitted only to the red side, as we
show in the lower panels of Figure~\ref{nb921_skew}.

\begin{figure*}[th]
\includegraphics[width=10cm,angle=0]{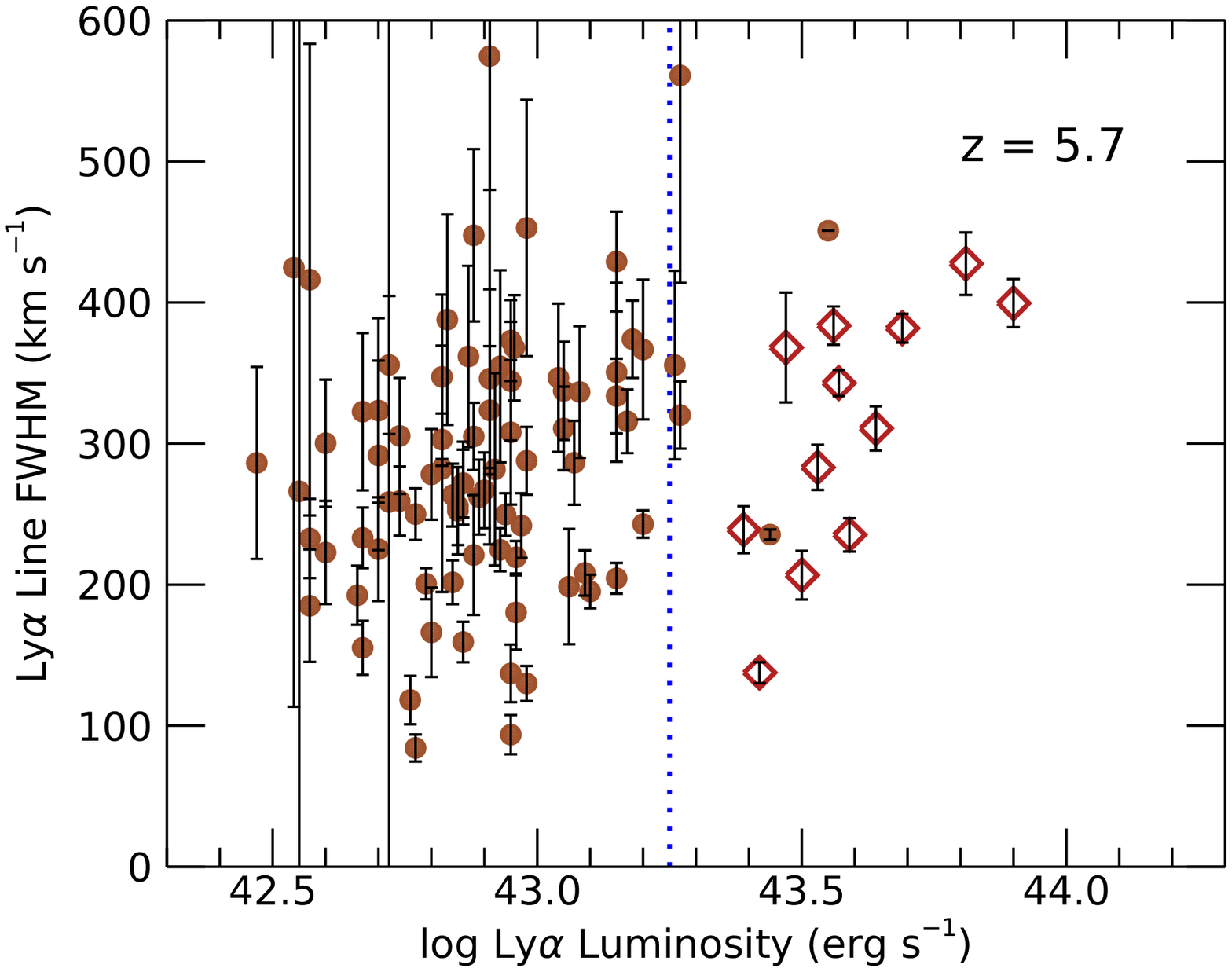}
\hspace{-1.0cm}
\includegraphics[width=10cm,angle=0]{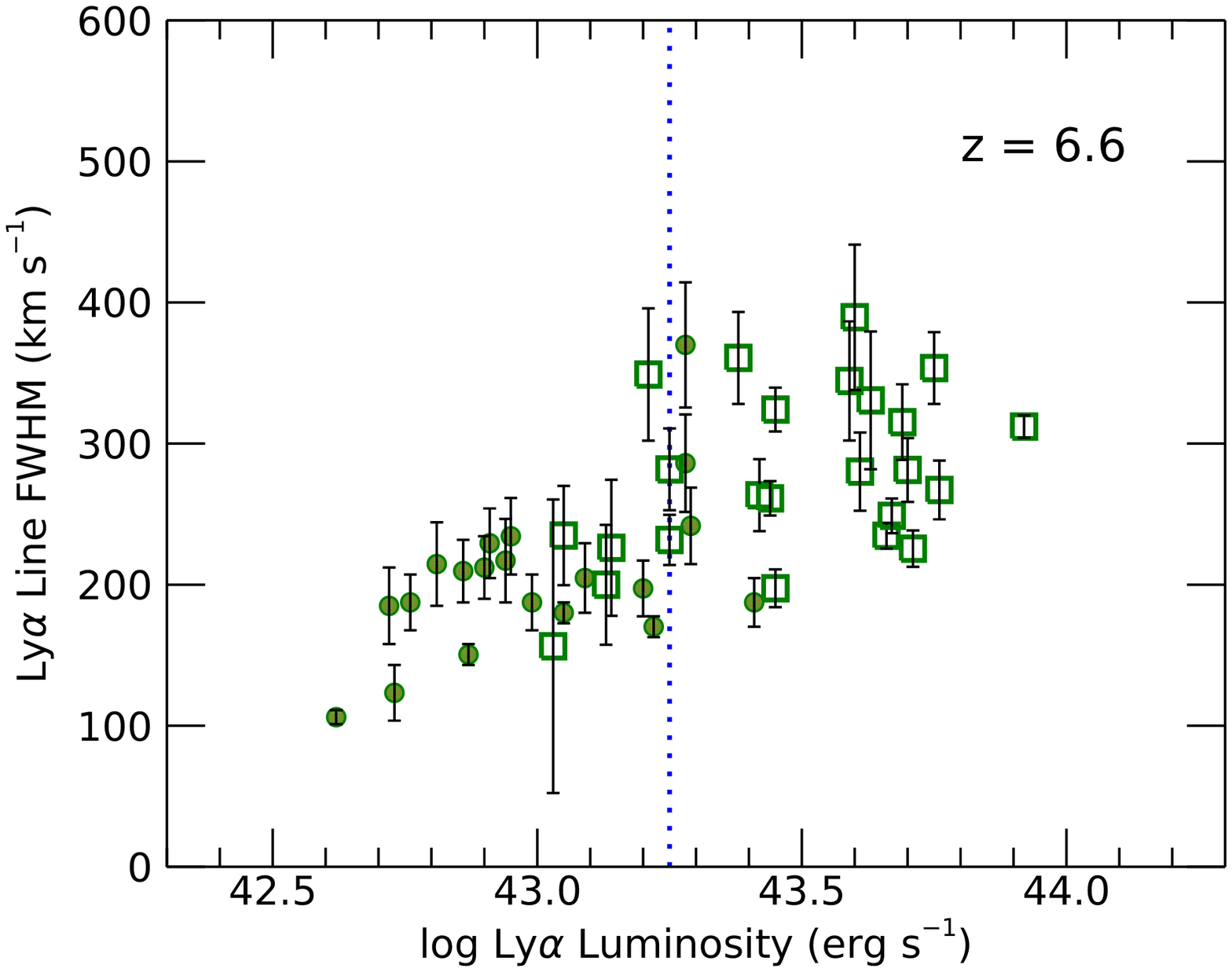}
\caption{Ly$\alpha$ line FWHM vs.\ $\log L (\textrm{\Lya})$
for the $z=5.7$ (left) data from Table~\ref{lumtab} (open diamonds)
and Hu et al.\ (2010) (solid circles), and for
the $z=6.6$ (right) data from Tables~\ref{lumtab2} and \ref{lumtab3} (open squares)
and Hu et al.\ (2010) (solid circles). The figure shows
the raw FWHM without a correction for the instrumental resolution.
In both panels, the blue dotted vertical line shows our rough division
of $\log L (\textrm{\Lya}) = 43.25$~erg~s$^{-1}$
between the lower- and higher-luminosity samples.
In all cases, the uncertainties are $1\sigma$. 
We do not show the very large error bars on the most luminous
Hu et al.\ (2010) object in the $z=5.7$ panel.
}
\label{widths_lums_z6_z5}
\end{figure*}

\begin{figure*}[th]
\hspace{-1.3cm}
\includegraphics[width=10cm,angle=0]{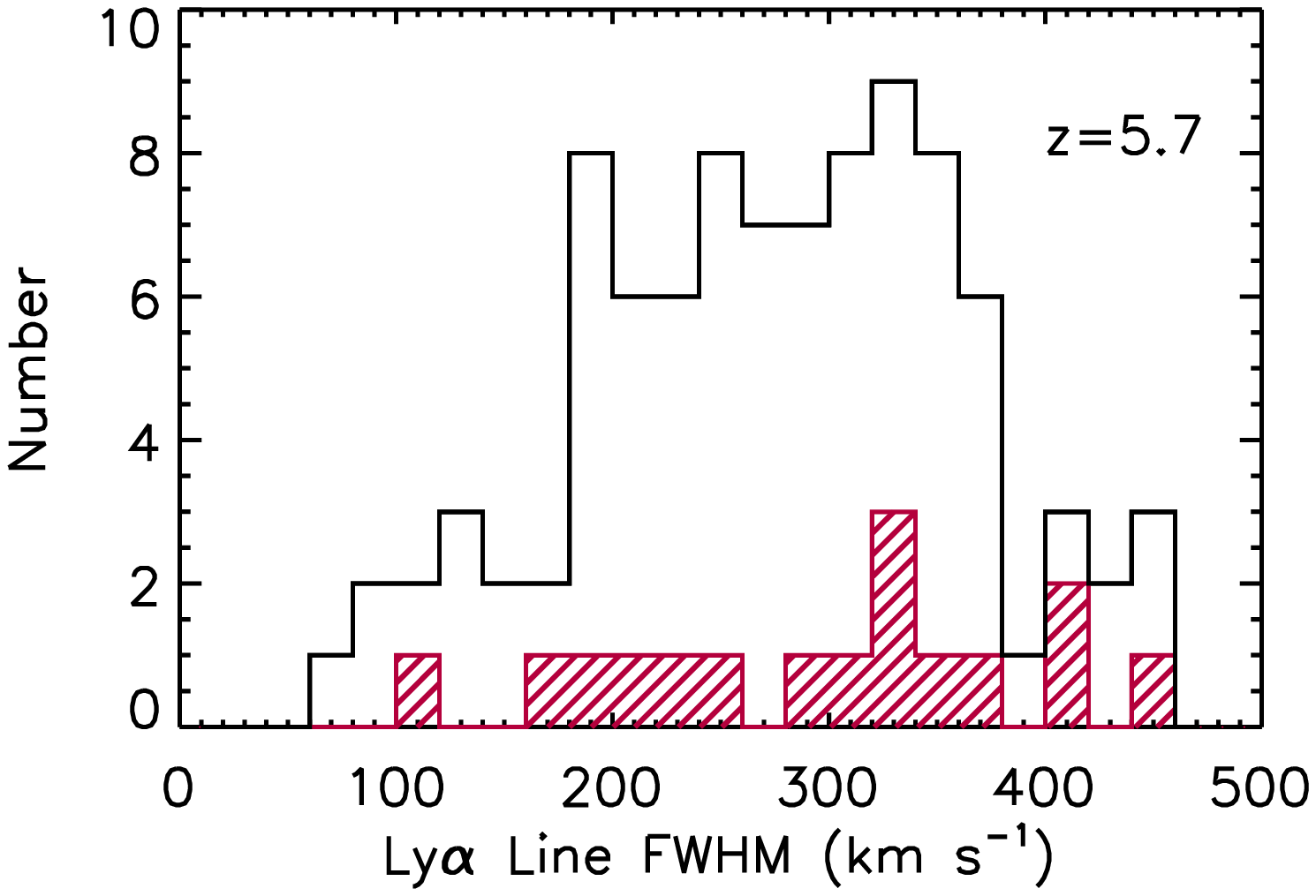}
\hspace{-1cm}
\includegraphics[width=10cm,angle=0]{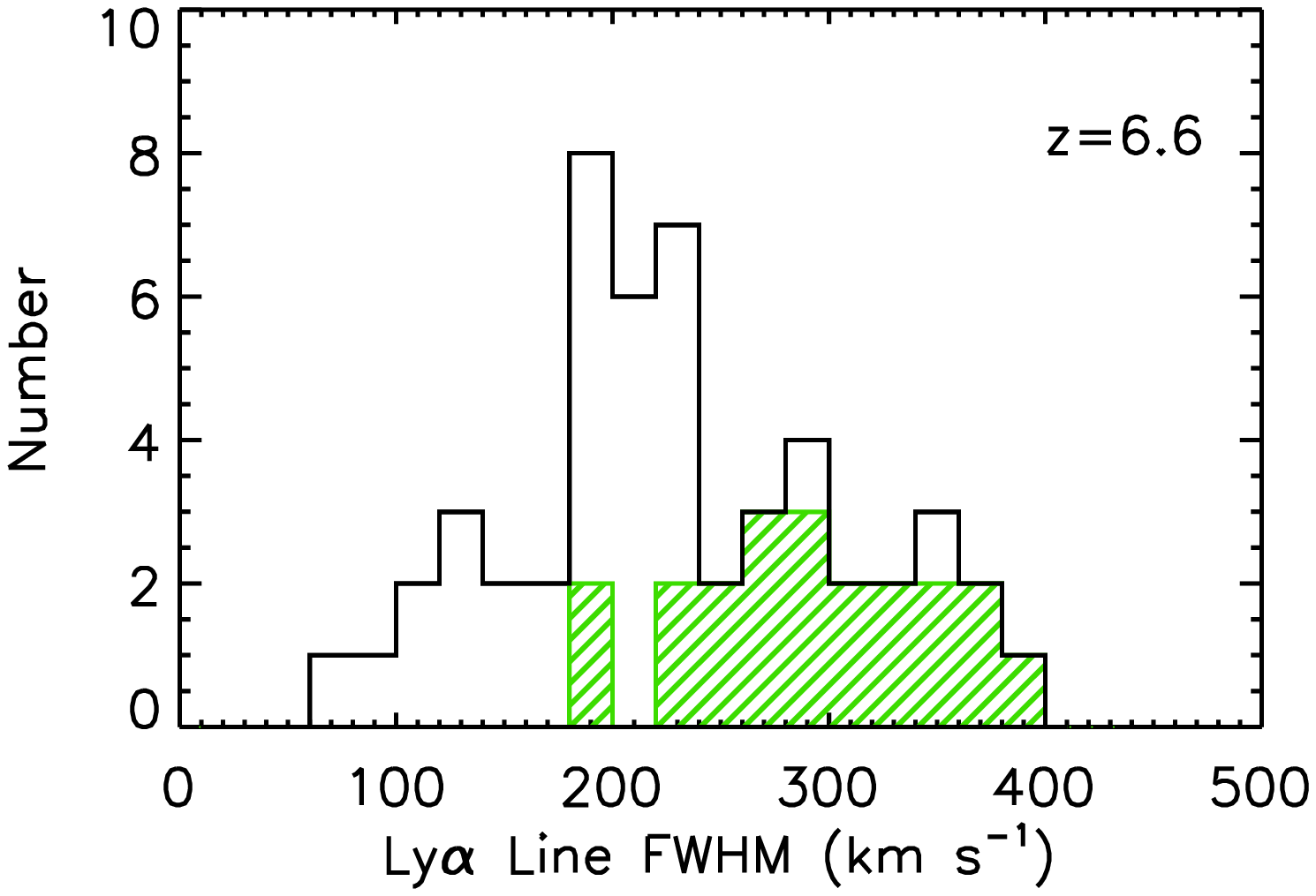}
\caption{Histograms of the distributions of the line widths
at $z=5.7$ (left) and $z=6.6$ (right) from Figure~\ref{widths_lums_z6_z5}.
In both panels, the open regions show the total
samples, and the shaded regions show the higher-luminosity samples. 
}
\label{hist_z6_z5}
\end{figure*}

For each LAE, we computed the FWHM of the
line and its error using the asymmetric profile
fit. We call this FWcalc, which has units of
km~s$^{-1}$. We give this line width in green in each panel of
Figures~\ref{nb921_skew} and \ref{nb816_skew}.
We show the fitted FWHM as the blue dashed line.
It is this quantity that we subsequently use in our analysis.
However, we also directly measured the FWHM from the
spectrum. The directly measured values are in broad agreement
with our adopted FWHMs, but they have a slight bias to lower values,
because they correspond to the first half-maximum
intercept. It is for this reason that we prefer the 
fitted FWHMs.

We give the fitted FWHMs and their 1$\sigma$ errors in the final columns
of Tables~\ref{lumtab}, \ref{lumtab2}, and \ref{lumtab3}. These values
are not corrected for the instrument resolution,
which has a FWHM of 117~km~s$^{-1}$. However, most
of the lines are very well resolved.

\section{Discussion}
\label{linewid}
In Figure~\ref{widths_lums_z6_z5},
we plot Ly$\alpha$ line FWHM versus $\log L (\textrm{\Lya})$ at $z=5.7$ (left) and
at $z=6.6$ (right).
The increase in the dynamical range of the measured Ly$\alpha$ luminosities
reveals a new result. At the previously measured lower luminosities, the 
widths of the lines show a decrease with increasing redshift,
but at the higher luminosities, the widths of the lines for the $z=5.7$ sample are 
comparable to the widths of the lines for the $z=6.6$ sample.

Based on the $z=6.6$ figure, we use a dividing line of 
$\log L (\textrm{\Lya}) = 43.25$~erg~s$^{-1}$ (blue dotted line)
to separate roughly the data into lower- and higher-luminosity samples. 
There is some uncertainty in this value, which could lie in the 43.17
to 43.4 range. We are currently expanding the sample of intermediate luminosity LAEs, 
which should allow a better determination. However, we note
that the analysis below is not sensitive to the exact choice.

Now we can quantify the result on the widths, which we do by redshift. At $z = 5.7$,
the median FWHM of the higher-luminosity sample ($\log L (\textrm{\Lya}) > 43.25$~erg~s$^{-1}$)
is fairly comparable to that of the lower-luminosity sample ($\log L (\textrm{\Lya}) <43.25$~erg~s$^{-1}$).
There are 17 sources in the higher-luminosity sample, which includes a very small number
of sources from \citet{hu10} (solid circles). 
The median FWHM for this sample is 320~km~s$^{-1}$,
with a standard error of 26~km~s$^{-1}$ computed using the bootstrap method.
For the 79 sources in the lower-luminosity sample,
which come entirely from \citet{hu10},
the median FWHM is 278~km~s$^{-1}$, with a 
standard error of 12~km~s$^{-1}$.

In contrast, at $z=6.6$, the median FWHM of the higher-luminosity sample is considerably larger
than that of the lower-luminosity sample. This is true for both the
sample of Tables~\ref{lumtab2} and \ref{lumtab3} (open squares) and the sample of \cite{hu10} 
(solid circles) in the rather limited overlap region.
For the 21 sources in the higher-luminosity sample, the median
FWHM is 281~km~s$^{-1}$, with a standard error of 21~km~s$^{-1}$. 
For the 30 $z=6.6$ sources in the lower-luminosity sample, the median FWHM is 
197~km~s$^{-1}$, with a standard error of 9~km~s$^{-1}$.

In Figure~\ref{hist_z6_z5}, 
we show the histograms of the distributions of the line widths at $z=5.7$ (left) and
$z=6.6$ (right) for the total (open regions) and higher-luminosity (shaded regions) 
samples of Figure~\ref{widths_lums_z6_z5}.
As expected, at $z=5.7$, the lower- and higher-luminosity samples have very similar distributions.
A Mann-Whitney test does not show a significant difference. 
However, at $z=6.6$, a Mann-Whitney test gives a probability of $< 10^{-5}$ that the lower-
and higher-luminosity samples have consistent distributions.

\begin{figure}[th]
\hspace{-0.5cm}
\includegraphics[width=10cm,angle=0]{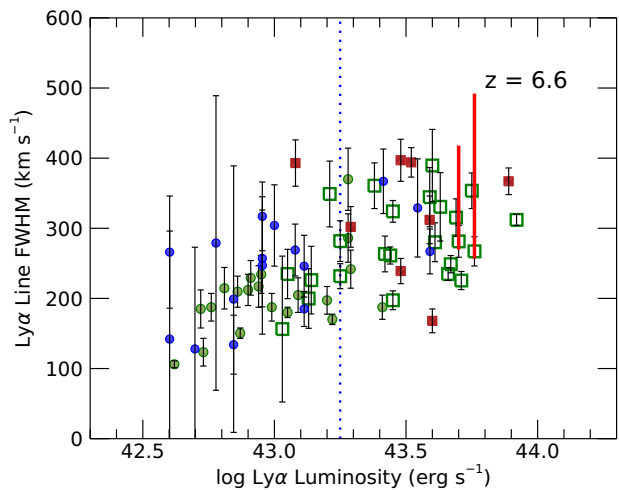}
\caption{Ly$\alpha$ line FWHM vs.\ $\log L (\textrm{\Lya})$ at $z=6.6$.
The green symbols of the present sample (from Figure~\ref{widths_lums_z6_z5})
are to be compared with the samples of 
\citet{shibuya18} (red squares) and \citet{ouchi10} (blue circles).
For the two double-peaked ULLAEs (COLA1 and NEPLA4, see Figure~\ref{nb921_skew}),
we show the range in FWHM when we include or exclude the blue wing (red bars).
}
\label{widths_lums_z6_shibuya}
\end{figure}

Moreover, including the blue wings for the two double-peaked 
sources COLA1 and NEPLA4 would only make this difference more pronounced, since it would
increase the higher-luminosity distribution. We illustrate this in
Figure~\ref{widths_lums_z6_shibuya}, where
we again plot Ly$\alpha$ line FWHM versus $\log L (\textrm{\Lya})$ for the $z=6.6$ samples
of Figure~\ref{widths_lums_z6_z5} (green symbols).
We use red vertical bars to show the range of FWHMs 
for COLA1 and NEPLA4 after excluding or including the blue wing.  

For comparison, we also show in this figure measurements 
from \cite{ouchi10} (blue) and \cite{shibuya18} (red).
Within the fairly substantial error bars on the \cite{ouchi10} 
data points, the present and archival data sets are broadly consistent. 
Note, however, that there is one archival lower-luminosity object 
\citep[HSCJ160107$+$550720 at $\log L (\textrm{\Lya}) = 43.08$~erg~s$^{-1}$ in ][]{shibuya18}
that has an unusually large width of 393~km~s$^{-1}$.

Because of the different instruments employed and the different analysis 
techniques used in these papers  versus the present work
(namely, single Gaussian versus asymmetric profile fitting),
we do not attempt to incorporate the \cite{ouchi10} and \cite{shibuya18}
 results into our analysis and instead
restrict to the current homogeneous samples of Tables 1--3.

\begin{figure*}[th]
\hspace{-0.5cm}
\includegraphics[width=9.75cm,angle=0]{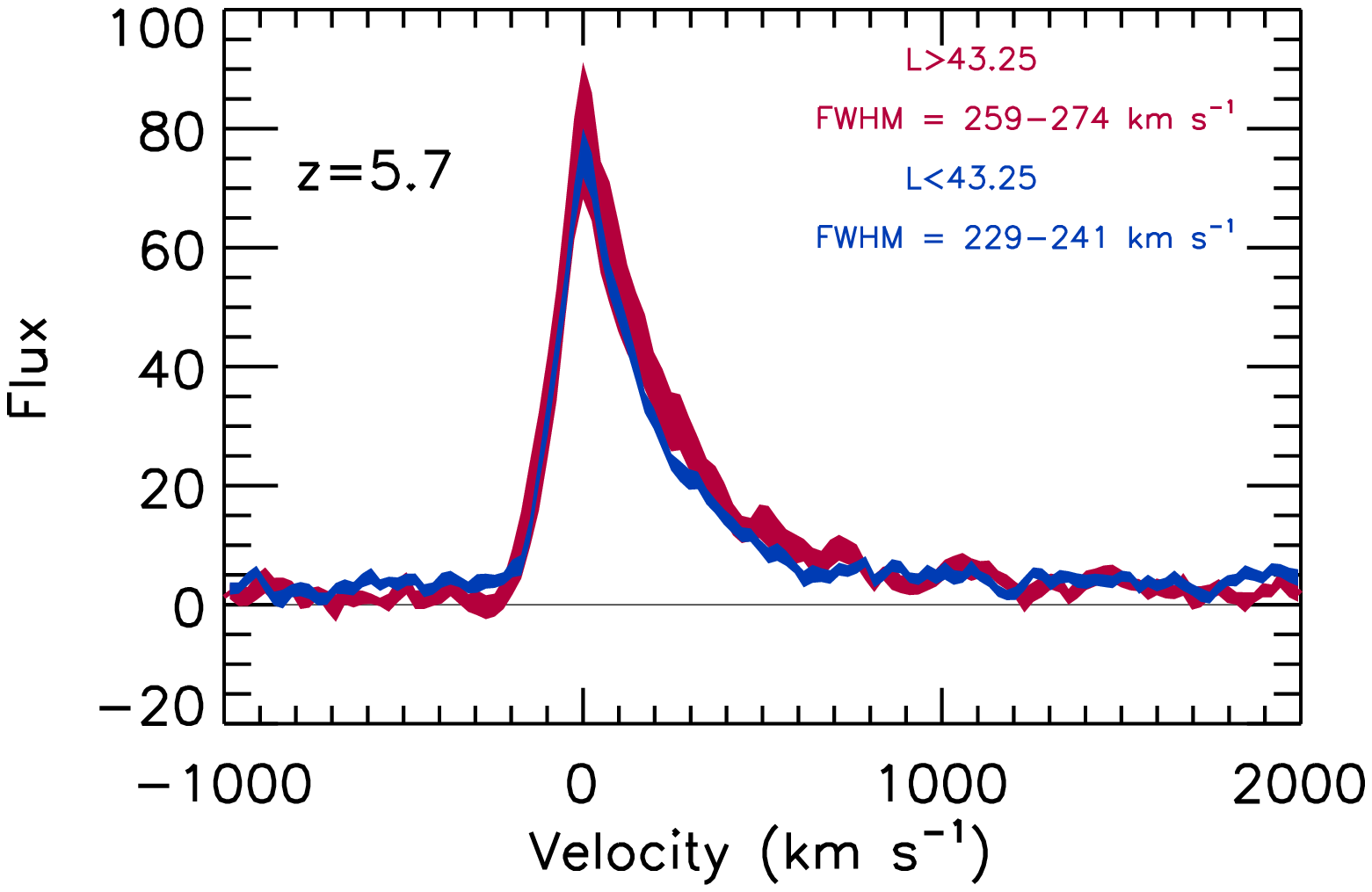}
\hspace{-1cm}
\includegraphics[width=9.75cm,angle=0]{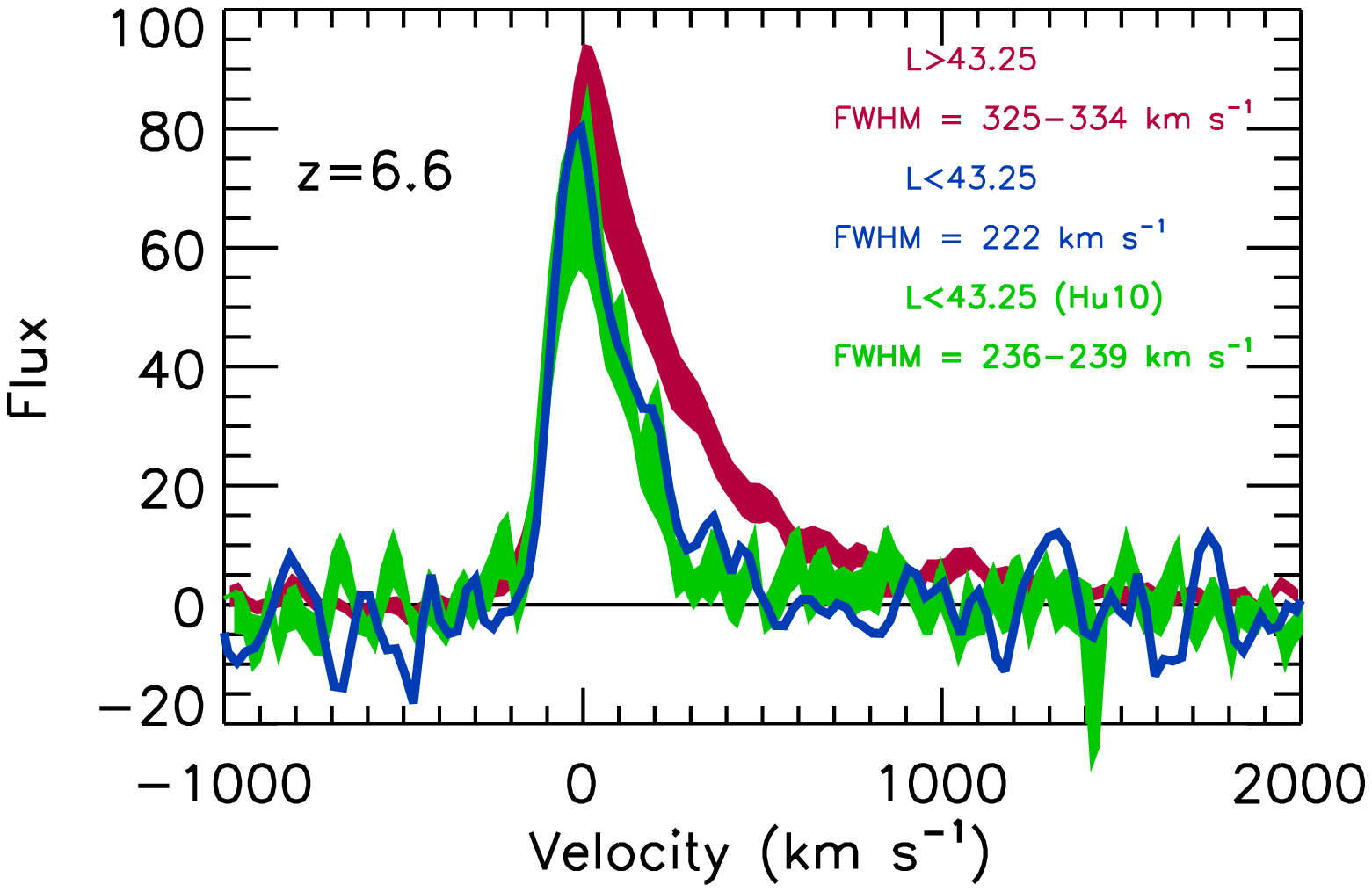}
\vskip -0.75cm
\caption{
{\em (Left)} Stack of the $z=5.7$ emission-line spectra for the 
lower-luminosity ($\log L (\textrm{\Lya}) < 43.25$~erg~s$^{-1}$; blue curve) 
LAEs from \cite{hu10} and
higher-luminosity ($\log L (\textrm{\Lya}) > 43.25$~erg~s$^{-1}$; red curve) 
LAEs of the present sample (Table~\ref{lumtab}).
{\em (Right)} Stack of the $z=6.6$ emission-line spectra for the 
lower-luminosity (blue curve) and higher-luminosity (red curve)  
LAEs of the present sample (Tables~\ref{lumtab2} and \ref{lumtab3}).
The green curve shows the stack for the lower-luminosity
LAEs from \citet{hu10}. (Note that the green and
blue curves are based on a disjoint set of objects.)
In both panels, the legends display 68\% confidence ranges for the FWHMs, 
which were computed using the bootstrap method.  
These confidence ranges are also displayed through shading.
In the right panel, the present $\log L (\textrm{\Lya}) < 43.25$~erg~s$^{-1}$ sample (blue curve)
has too few objects to compute the error range.
}
\label{stack_z6_z5}
\end{figure*}

We can compare the lower- and higher-luminosity 
samples more directly by stacking the individual spectra.
In Figure~\ref{stack_z6_z5}, we show the sum of the spectra at $z=5.7$ (left) and 
at $z=6.6$ (right) for the various samples.
The shading shows the 68\% confidence range calculated using the bootstrap method.
The $z=5.7$ stacked spectra show close
agreement between the lower- and higher-luminosity samples, with the higher-luminosity 
sample being only slightly wider, while the $z=6.6$ stacked spectra show 
the higher-luminosity sample being considerably wider.
In all cases, the measured FWHMs of the stacks match 
well to the median FWHMs of the individual sources discussed above.

\begin{figure*}
\center{\includegraphics[width=9.5cm,angle=0]{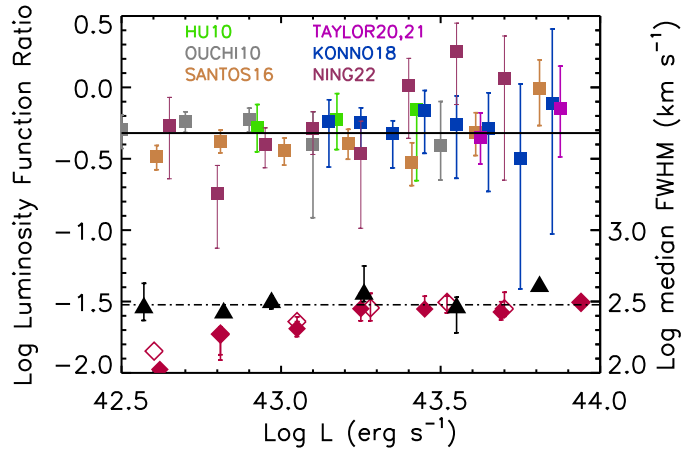}}
\vskip -0.75cm
\caption{Logarithmic ratio of the LF at $z=6.6$ to that at $z=5.7$
for all published LAE LFs (solid squares color-coded according to the legend).
The black solid line corresponds to a drop of 2.1
between the two redshifts, which roughly matches
all of the measurements at lower luminosities.
At the high-luminosity end, the measurements
show a logarithmic ratio closer to 1, but the results are only marginally
significant. The lower points (right-axis scale)
show the median FWHMs at $z=6.6$ (red solid diamonds are the present data,
while the red open diamonds
also include the \citealt{shibuya18} and \citealt{ouchi10} data) 
and at $z=5.7$ (black solid triangles). The black dot-dashed line shows a value of
300~km~s$^{-1}$, which roughly matches the entire $z=5.7$ sample
and the $z=6.6$ higher-luminosity sample. All error bars
are 68\% confidence ranges.
}
\vskip 1cm
\label{plot_lfs}
\end{figure*}

\section{Summary}
\label{secphys}

We summarize our results in Figure~\ref{plot_lfs},
where we compare the evolution of the LFs with the evolution
of the line widths. The solid squares show the ratio
of the LF at $z=6.6$ to that at $z=5.7$ as reported by all of the
groups who have measured the two LFs \citep{hu10,ouchi10,santos16,
konno18,taylor20,taylor21,ning22}. While there is substantial variation
in the normalization of the LFs between the groups, the LF ratio
is much more homogeneous. All of the measurements show a drop
of about 2.1 in the $z=6.6$ LF from that at $z=5.7$. Only at
the high-luminosity end is there a smaller drop, but although 
this result is seen in all the samples, it is somewhat marginal
(see \citealt{taylor21} for a more extended discussion).

By contrast, the evolution of the line widths is much more clear.
The median FWHMs as a function of $\log L (\textrm{\Lya})$ are shown
as the black triangles ($z=5.7$) and the red diamonds ($z=6.6$).
The solid diamonds include only the present data,
while the open diamonds also include the data from
\citet{ouchi10} and \citet{shibuya18}. 
The $z=5.7$ values show
no variation with luminosity, consistent with their
lying in similarly ionized regions of the IGM. 
The $z=6.6$ values at higher luminosities
are $\sim 300$~km~s$^{-1}$, the same 
as the $z=5.7$ values at all luminosities.
However, the $z=6.6$ values at lower luminosities show
a highly significant drop, consistent with the higher-redshift
sources lying in more neutral IGM.

We conclude that $z=6.6$ LAEs with observed luminosities
$\log L (\textrm{\Lya}) > 43.25~{\rm erg\ s}^{-1}$ mark
ionized regions. The reason for this could be that the galaxies
themselves are fully responsible for the ionization,
but it is also possible that they are lying in regions
where neighboring galaxies are producing sufficient
ionization to allow the intrinsic galaxy profile to be seen. 

While this seems the most likely explanation, there may be
other possibilities. In particular, intrinsic line profile shapes may vary
due to different physical conditions, dust extinction, or the 
age of the star formation burst
(see, e.g., \citealt{verhamme12,verhamme15,naidu22}), and this
could result in differential evolution between galaxies of 
different luminosities.

If the luminosity above where there are ionized bubbles is 
$\log L (\textrm{\Lya}) = 43.25~{\rm erg\ s}^{-1}$ 
rather than $43.5~{\rm erg\ s}^{-1}$, then this substantially increases
the comoving number density from 
$7\times10^{-7}$~Mpc$^{-3}$ to $4\times10^{-6}$~Mpc$^{-3}$ based
on the LFs of \citet{taylor20,taylor21}. If we adopt a minimum
bubble radius of around 4~Mpc to allow for Ly$\alpha$ escape
\citep{meyer20,gronke21}, then the higher value would correspond
to a filling factor greater than 1\% for the bubbles.
However, the value could be substantially larger if the bubbles
have larger radii.
This is consistent with the idea that the most luminous sources 
are driving the reionization (see, e.g., \citealt{matthee22}).

The next steps will include, first, increasing the number of measured
FWHMs near the transition luminosity to see how
abrupt any transition is, and, second, searching for objects neighboring 
the higher-luminosity LAEs to characterize the
ionization states of the regions. Finally, we aim to improve
the determination of the LFs for the higher-luminosity LAEs.

\begin{acknowledgements}
We thank the anonymous referee for a very constructive report that helped us to improve 
the manuscript.
We gratefully acknowledge support for this research from NSF grants AST-1716093 
(E.~M.~H., A.~S.) and AST-1715145 (A.~J.~B). We also gratefully acknowledge the
William F. Vilas Estate (A.~J.~B.), a Kellett Mid-Career Award and a WARF Named 
Professorship from the University of Wisconsin-Madison
Office of the Vice Chancellor for Research and Graduate Education
with funding from the Wisconsin Alumni Research Foundation (A.~J.~B.),
a WSGC Graduate and Professional Research Fellowship (A.~J.~T.), and
a Sigma Xi Grant in Aid of Research (A.~J.~T.).

The W.~M.~Keck Observatory is operated as 
a scientific partnership among the California Institute of Technology, the 
University of California, and NASA, and was made possible by the generous financial 
support of the W.~M.~Keck Foundation.

We wish to recognize and acknowledge the very significant 
cultural role and reverence that the summit of Maunakea has always 
had within the indigenous Hawaiian community. We are most fortunate 
to have the opportunity to conduct observations from this mountain.
\end{acknowledgements}

\facilities{Keck}

\bibliography{Lyabib.bib}

\begin{deluxetable*}{lrrrcr}
\tablecaption{HEROES $z=5.7$\ LAE Ly$\alpha$ Line Luminosities and Widths \label{lumtab}}
\tablewidth{350pt}
\tablehead{
\colhead{Name} & \colhead{R.A.} & \colhead{Decl.} & 
\colhead{$z$} &  
\colhead{$\log L({\rm Ly}\alpha$)} & \colhead{FWHM} \\
\colhead{} & \colhead{} & \colhead{} & \colhead{} & \colhead{} &  
\colhead{${\rm km\ s}^{-1}$} 
}
\startdata 
NEPLA271.34$+$67.92 & 271.34009 & 67.92042 & 5.719  & 43.90 & $399.5 \pm 17.1$ \cr
NEPLA267.89$+$66.63 & 267.89209 & 66.62925 & 5.696  & 43.81 & $427.5 \pm 22.2$ \cr
NEPLA262.36$+$68.03 & 262.36445 &   68.03414 &   5.738 &   43.64 & $310.7 \pm 15.7$ \cr
NEPLA272.58$+$66.69 &   272.58476 &   66.69040 &   5.738 &  43.69 & $381.8 \pm 10.2$\cr
NEPLA267.79$+$66.73 &   267.78810 &   66.72774 &   5.695 &  43.59 & $235.3 \pm 11.8$ \cr
NEPLA274.55$+$67.85 &   274.55319 &   67.84963 &   5.722 &  43.57 & $342.9 \pm 9.27$ \cr
NEPLA263.36$+$68.20 &   263.36466 &   68.19836 &   5.696 &  43.56 & $383.6 \pm 13.6$ \cr
NEPLA265.64$+$68.50 &   265.63687 &   68.49751 &   5.718 &  43.50 & $206.7 \pm 17.2$ \cr
NEPLA275.87$+$64.57 &   275.87088 &   64.56588 &   5.732 &  43.53 & $283.1 \pm 16.0$\cr
NEPLA263.44$+$66.33 &   263.44406 &   66.33290 &   5.714 &  43.47 & $368.1 \pm 38.9$ \cr
NEPLA272.36$+$64.83 &   272.35809 &   64.83355 &   5.699 &  43.42 & $137.6 \pm 7.43$\cr
NEPLA271.72+64.86 & 271.72323  &  64.85677 &   5.703 &  43.39  & $238.9 \pm 16.6$
\enddata
\end{deluxetable*}

\begin{deluxetable*}{lrrrcr}
\tablecaption{HEROES and Other $z=6.6$\ LAE Ly$\alpha$ Line Luminosities and Widths \label{lumtab2}}
\tablewidth{350pt}
\tablehead{
\colhead{Name} & \colhead{R.A.} & \colhead{Decl.} & 
\colhead{$z$}  & 
\colhead{$\log {L(\rm Ly}\alpha$)} & \colhead{FWHM} \\
\colhead{} & \colhead{} &  \colhead{} & \colhead{} & \colhead{} &  
\colhead{${\rm km\ s}^{-1}$} 
}
\startdata 
COLA1 & 150.64751 & 2.20375 & 6.5923  &  43.70 &  $281.3 \pm 22.6$ \cr
CR7 & 150.24167 & 1.80422 & 6.6010 &  43.67 & $248.8 \pm 12.3$ \cr
MASOSA & 150.35333 & 2.52925 & 6.5455  &  43.42 & $263.4 \pm 25.5$ \cr
GN-LA1 & 189.35817 & 62.20769 & 6.5578 &  43.45 & $197.4 \pm 13.4$ \cr
NEPLA1 & 273.73837 & 65.28599 & 6.5938 &  43.92 &  $312.0 \pm 7.95$ \cr
NEPLA2 & 263.61490 & 67.59397 & 6.5831 & 43.71 & $225.5 \pm 12.9$  \cr
NEPLA3 & 265.22437 & 65.51036 & 6.5915 & 43.66 & $234.7 \pm 9.06$  \cr
NEPLA4 & 268.29211 & 65.10958 & 6.5472 & 43.76 & $267.1 \pm 20.8$ \cr
NEPLA5 & 269.68964 & 65.94475 & 6.5364 & 43.60 & $389.5 \pm 51.5$ \cr
NEPLA6 & 262.44296 & 65.18044 & 6.5660 & 43.75 & $353.4 \pm 25.4$ \cr
NEPLA7 & 272.66104 & 67.38605 & 6.5780&43.59 & $344.4 \pm 42.2$ \cr
NEPLA8 & 262.30838 & 65.59966 & 6.5668 & 43.61 & $280.1 \pm 27.7$ \cr
NEPLA9 &276.23441 & 67.60667 &  6.5352 & 43.63 & $330.6 \pm 48.8$ \cr
VR7 & 334.73483 &  0.13536 & 6.5330 & 43.69 & $315.2 \pm 26.8$
\enddata
\end{deluxetable*}

\begin{deluxetable*}{lrrrcr}
\tablecaption{JTDF $z=6.6$\ LAE Ly$\alpha$ Line Luminosities and Widths \label{lumtab3}}
\tablewidth{350pt}
\tablehead{
\colhead{Name} & \colhead{R.A.} & \colhead{Decl.} & \colhead{$z$} &  
\colhead{$\log{L(\rm Ly}\alpha$)}  & \colhead{FWHM} \\
\colhead{} & \colhead{} & \colhead{} & \colhead{} & \colhead{} & 
\colhead{${\rm km\ s}^{-1}$} 
}
\startdata 
  NEPLA259.78+65.38  &  259.78860  &  65.38805  &   6.5750   &  43.25 & $281.7 \pm 29.0$ \cr
  NEPLA259.91+65.38 &  259.91315  &  65.38100  &   6.5750   &  43.44 & $261.2 \pm 12.2$ \cr
  NEPLA260.66+66.13  &  260.66571 &   66.13014 &    6.5785   &  43.05 &  $234.8 \pm 35.2$ \cr
  NEPLA260.79+66.06   &  260.79163  &  66.06917  &   6.5938   & 43.25 & $231.7 \pm 17.8$ \cr 
  NEPLA260.80+65.37   &  260.80258  &  65.37336   &  6.5808   &  43.45 & $324.1 \pm 15.5$ \cr
  NEPLA260.88+65.40   &  260.88062  &  65.40775  &   6.5463   &  43.13 & $199.9 \pm 42.5$ \cr
  NEPLA261.67+65.83   &  261.67025  &  65.83728  &   6.5492   &  43.14 & $226.1 \pm 48.2$ \cr
  NEPLA261.70+65.79  &  261.70859  &  65.79608  &   6.5615   & 43.38 & $360.7 \pm 32.6$ \cr
  NEPLA262.02+66.04   &  262.02902  &  66.04408  &   6.5990   &  43.21 & $348.9 \pm 46.9$\cr 
  NEPLA262.46+65.66  &  262.46057 &   65.66861  &   6.5665   &  43.03 &  $156.3 \pm 104.$ 
\enddata
\end{deluxetable*}

\end{document}